\documentclass[prd,aps,twocolumn,showpacs,preprintnumbers,amsmath,amssymb,nofootinbib]{revtex4}
\usepackage[dvips]{graphicx}
\usepackage{epsf}
\usepackage{amsmath}
\usepackage{amssymb}

\usepackage{graphicx}
\usepackage{dcolumn}
\usepackage{bm}
\begin{document}

\title{Modeling $f(R)$ gravity in terms of mass dilation rate }

\author{Jian-hua He$^{1}$\email{jianhua.he@brera.inaf.it}, Bin Wang$^{2}$}

\affiliation{$^{1}$ INAF-Osservatorio Astronomico
di Brera, Via Emilio Bianchi, 46, I-23807, Merate
(LC), Italy}
\affiliation{$^{2}$ INPAC and
Department of Physics, Shanghai Jiao Tong
University, Shanghai 200240, China}


\begin{abstract}
We review the conformal equivalence in
describing the background expansion of the
universe by $f(R)$ gravity both in the Jordan
frame and the Einstein frame. In the Jordan
frame, we present the general analytic expression
for $f(R)$ models that have the same expansion
history as the $\Lambda$CDM model. This analytic
form can provide further insights on how
cosmology can be used to test the $f(R)$ gravity
at the largest scales. Moreover we present a
systematic and self-consistent way to construct
the viable $f(R)$ model in Jordan frame using the
mass dilation rate function from the Einstein
frame through the conformal transformation. In
addition, we extend our study to the linear
perturbation theories and we further exhibit the
equivalence of the $f(R)$ gravity presented in
the Jordan frame and Einstein frame in the
perturbed space-time. We argue that this
equivalence has solid physics root.

\end{abstract}

\maketitle

\section{Introduction}
There has been conclusive evidence indicating
that our universe is experiencing an accelerated
expansion. This acceleration is believed driven
by a so called dark energy (DE) in the framework
of Einstein's general relativity. The simplest
explanation of such  DE is the cosmological
constant. But it suffers serious problems such
that its value is far below the prediction of any
sensible quantum field theories and it inevitably
sustains the coincidence problem with the
mysterious same order energy density as the
matter field today.

There exists an alternative way to explain the
acceleration of the universe expansion by
modifying the Einstein's gravity.  One simplest
attempt is called $f(R)$ gravity, in which the
scalar curvature in the Lagrange density of
Einstein's gravity is replaced by an arbitrary
function of $R$ \cite{fr}. However it is quite
non-trivial to construct the viable $f(R)$ model
which can satisfy both cosmological and local
gravity
constraints~\cite{34,35,36,37,38,39,40,AmendolaPRL,Amendolafrcondition}.
Only a few viable $f(R)$ models with simple
analytic forms have been found
~\cite{Amendolafrcondition,wayneHu,staobinsky,Tsujikawa}.
However, if we do not limit to these $f(R)$ models with
explicit expressions, in a more general case,
$f(R)$ gravity in cosmology can leave enough
freedoms to accommodate any desired expansion
histories of the universe
~\cite{solution}~\cite{Song}. A general
phenomenological $f(R)$ cosmological model were
formulated numerically  by specifying the
effective DE equation of state $w$ and the
present value of the parameter $B\equiv\frac{d\ln
F}{d\ln H}$~\cite{Song,Hu}.

On the other hand, it is always possible to
transform the theory of $f(R)$ gravity presented
in the Jordan frame to the Einstein frame by
using the conformal transformation \cite{dicke}.
In the Einstein frame, the model contains the
coupling between the matter and an additional
scalar field \cite{tsujikawa}. If the scalar
field plays a role of DE in the universe, DE
would interact with DM. The phenomenological DE
and DM interacting model has been studied
extensively in literatures, see
\cite{7,HJH,Hepertur,coupling,199,1,2} and
references therein. More interestingly, it was
found that a general  $f(R)$ gravity  can be
consistently constructed in a covariant form in
terms of the mass dilation rate function in the
Einstein frame\cite{Hefr}. This mass dilation
rate function  marks the coupling strength
between DE and DM. The condition for the $f(R)$
cosmology to avoid the instability at
high-curvature region and be consistent with CMB
observation require the energy flow from DE to DM
in the framework of the interacting model
\cite{Hefr}. This was found helpful to alleviate
the coincidence problem\cite{coupling,199}.

In this paper, we will further examine the
equivalence of the $f(R)$ gravity presented in
the Jordan frame and the Einstein frame to
describe the  cosmology. In the background Jordan
frame, we will derive the analytic $f(R)$
functions to give the same expansion history of
the universe as the $\Lambda$CDM model. This can
provide further insights on how cosmology can be
used to test the $f(R)$ gravity at the largest
scales. In addition to modeling $f(R)$ gravity in
Jordan frame using the effective DE equation of
state, we will present a systematic and
self-consistent way to construct the $f(R)$ model
in terms of the mass dilation rate function in
Einstein frame. Moreover we will develop linear
perturbation theories to further exhibit the
equivalence of the $f(R)$ gravity in the Jordan
frame and the Einstein frame.

This paper is orgnized as follows: In
section~\ref{background}, we will discuss the
$f(R)$ gravity in the background cosmology. In
the Jordan frame, we will derive the general
analytical solutions for the $f(R)$ model which
has the same expansion history as the
$\Lambda$CDM cosmology. In the Einstein frame, we
will argue that once the mass dilation is
specified, the $f(R)$ gravity can be constructed
accordingly.  In
section~\ref{sectionperturbedspacetime}, we will
extend our discussion to the perturbed
space-time. We will show that perturbation
theories in Jordan frame and Einstein frame can
be consistently connected by conformal
transformations. In section~\ref{subhoizion}, we
will construct a viable $f(R)$ cosmological model
in terms of mass dilation rate specified in the
Einstein frame and discuss the subhorizon
approximation of the constructed $f(R)$ model.
Our conclusion and discussion will be presented
in the last section.

\section{Background dynamics in $f(R)$ cosmology\label{background}}
We start with the 4-dimensional action in $f(R)$ gravity~\cite{staobinskyfr}
\begin{equation}
S=\frac{1}{2\kappa^2}\int d^4x\sqrt{-g}f(R)+\int d^4x\mathcal{L}^{(m)}\quad,
\end{equation}
where $\kappa^2=8\pi G$. The field equation can
be obtained by varying the above action with
respect to $g_{\mu\nu}$, which leads to
\begin{equation}
FR_{\mu\nu}-\frac{1}{2}fg_{\mu\nu}-\nabla_{\mu}\nabla_{\nu}F+g_{\mu\nu}\Box F=\kappa^2T_{\mu\nu}^{(m)}\quad ,\label{FRfield}
\end{equation}
where $F=\frac{\partial f}{\partial R}$ and
\begin{equation}
\Box F=\frac{1}{\sqrt{-g}}\partial_{\mu}(\sqrt{-g}g^{\mu\nu}\partial_{\mu}F)\quad.
\end{equation}
If we define the left hand side of Eq.~(\ref{FRfield}) as a tensor $\Sigma_{\mu\nu}$
\begin{equation}
\Sigma_{\mu\nu}=FR_{\mu\nu}-\frac{1}{2}fg_{\mu\nu}-\nabla_{\mu}\nabla_{\nu}F+g_{\mu\nu}\Box F\quad ,\label{MDEinstein}
\end{equation}
we can recast the modified Einstein's equation
into a similar form to the standard Einstein
equation
\begin{equation}
\Sigma_{\mu\nu}=\kappa^2T_{\mu\nu}^{(m)}\label{Einsteineq}\quad.
\end{equation}
The trace of the above equation gives rise to a
scalar equation
\begin{equation}
f=\frac{3}{2}\Box F + \frac{1}{2}FR-\frac{\kappa^2}{2} T^{(m)}\quad .
\end{equation}
Taking the derivative of $f$ and further noting that
\begin{equation}
\nabla_{\mu}f = F\nabla_{\mu}R \quad,
\end{equation}
we obtain
\begin{equation}
\nabla_{\mu}R =-\frac{\kappa^2}{F}\nabla_{\mu}T^{(m)}+\frac{3}{F}\nabla_{\mu}\Box F + R\nabla_{\mu}\ln F\quad .
\end{equation}
We can see that $F$ is a free scalar, which can
be used to characterize the $f(R)$ gravity.
\subsection{The Jordan frame}
In the Jordan frame, we consider the universe
which is described by the flat
Friedmann-Robertson- Walker(FRW) metric
\begin{equation}
ds^2 =-dt^2+a^2dx^2.
\end{equation}
From Eq.(\ref{Einsteineq}), we can obtain the
equation governing the dynamics in the background
spacetime
\begin{equation}
H^2=\frac{FR-f}{6F}-H\frac{\dot{F}}{F}+\frac{\kappa^2}{3F}\rho_m\label{field}\quad.
\end{equation}
In the Jordan frame, the matter field is
conserved, which satisfies
\begin{equation}
\dot{\rho}_m+3H\rho_m=0\quad,
\end{equation}
where dot denotes the derivative with respect to
the cosmic time $t$.

From Eq.(\ref{field}), $f$ can be presented as
\begin{equation}
f=FR-6FH^2-6H^2\frac{dF}{dx}+2\kappa^2\rho_m\label{fHubble}\quad.
\end{equation}
where $x=\ln a$.

Taking the derivative of the above equation and
further noting that $\frac{df}{dx}=F\frac{d
R}{dx}$, we obtain
\begin{equation}
\frac{d^2}{dx^2}F+(\frac{1}{2}\frac{d\ln E}{dx}-1)\frac{dF}{dx}+\frac{d\ln E}{dx}F+\frac{3\Omega_m^0e^{-3x}}{E}=0\label{F}\quad,
\end{equation}
where
\begin{eqnarray}
E&\equiv&\frac{H^2}{H_0^2}\nonumber\quad, \\
\Omega_m^0&\equiv&\frac{\kappa^2\rho_m^0}{3H_0^2}\quad,\\
R&=&3(\frac{dE}{dx}+4E)\quad.
\end{eqnarray}
$R$ is in the unit of $H_0^2$ and $\kappa^2=1$.

For illustrative purposes, we take an expansion
history in the Jordan frame that matches a DE
model with equation of state $w$,
\begin{equation}
E=\Omega_m^0e^{-3x}+(1-\Omega_m^0)e^{-3\int_0^x(1+w)dx}\label{history}\quad.
\end{equation}
In \cite{Song} a family of numerical solutions of
$f(R)$ models that match the representative
expansion histories of DE model was obtained.
Here we find that for the $f(R)$ model to mimic
the same expansion history as that of the
$\Lambda$CDM cosmology in the background $E=
\Omega_m^0e^{-3x}+(1-\Omega_m^0)\quad$,  we can
analytically solve Eq.(\ref{F}) and obtain the
general analytical solution
\begin{eqnarray}
&&F(x)-1= \nonumber\\
&&C(e^{3x})^{p_-}(1-\Omega_m^0)^{p_-}{(\Omega_m^0)}^{-p_-}{_2F_1}[q_-,p_-,r_-,z]\nonumber\\
&&+D(e^{3x})^{p_+}(1-\Omega_m^0)^{p_+}{(\Omega_m^0)}^{-p_+}{_2F_1}[q_+,p_+,r_+,z] \nonumber\quad,\\\label{solution}
\end{eqnarray}
where $_2F_1(a,b;c;z)$ is the hypergeometric
function, which is defined as
\begin{eqnarray}
_2F_1(a,b;c;z)&=&\sum_{n=0}^{\infty}\frac{(a)_n(b)_n}{(c)_n}\frac{z^n}{n!}\quad ,\nonumber\\
(a)_n&=&a(a+1)\cdots(a+n-1) \nonumber \quad .
\end{eqnarray}
$C$ and $D$ are coefficients which will be
determined by boundary conditions. The indexes in
the solution read,
\begin{eqnarray}
p_+&=&\frac{5+\sqrt{73}}{12}\nonumber \quad,\\
p_-&=&\frac{5-\sqrt{73}}{12}\nonumber \quad,\\
q_+&=&\frac{1+\sqrt{73}}{12}\nonumber \quad,\\
q_-&=&\frac{1-\sqrt{73}}{12}\nonumber \quad,\\
r_+&=&1+\frac{\sqrt{73}}{6}\nonumber \quad,\\
r_-&=&1-\frac{\sqrt{73}}{6}\nonumber\quad.\\
\end{eqnarray}
and
\begin{eqnarray}
z&=&e^{3x}(1-\frac{1}{\Omega_m^0})\nonumber \quad,\\
e^{3x}&=&\frac{3\Omega_m^0}{R-12(1-\Omega_m^0)}\quad.
\end{eqnarray}

Since the index $p_-$ is negative, the fist term
of Eq.(\ref{solution}) will diverge when $x$ goes
to the minus infinity, which is not acceptable in
physics. Besides, a viable $f(R)$ model should be
in a form of ``chameleon" type ~\cite{Khoury},
which means that in the high curvature region it
should go back to the standard Einstein gravity
to pass the local Solar-System tests. The
boundary condition for the scalar field $F(x)$
thus requires
\begin{equation}
\lim_{x\rightarrow -\infty} F(x) = 1\quad,\label{boundary}
\end{equation}
and we need to set $C=0$.  The solution turns out
to be
\begin{equation}
F(x)=1+D(e^{3x})^{p_+}(1-\Omega_m^0)^{p_+}{(\Omega_m^0)}^{-p_+}{_2F_1}[q_+,p_+,r_+,z]\label{Fsolution}\quad.
\end{equation}
$D$ above is a free parameter which characterizes
the different $f(R)$ models which have the same
expansion history as that of the $\Lambda$CDM
model.

With the solution of $F(x)$, we can easily figure
out the analytical expression for $f(R)$ by doing
the integration
\begin{equation}
f(R)=f(R(z))=\int F(z) \frac{\partial R}{\partial z}dz\label{integral}\quad .
\end{equation}
We obtain
\begin{equation}
f(R)=[L_{\Lambda}-6(1-\Omega_m^0)](1-\epsilon)+{\rm Constant}\label{finallfR}\quad ,
\end{equation}
where
\begin{equation}
\epsilon=D\frac{\Gamma_{E}[s_+]}{\Gamma_{E}[p_+]}(1-\Omega_m^0)^{p_+}(\Omega_m^0)^{-p_+}(e^{3x})^{p_+}{_2F_1}[q_+,s_+,r_+,z]\nonumber\quad,
\end{equation}
and
\begin{eqnarray}
L_{\Lambda}&=&R-2\Lambda=R-6(1-\Omega_m^0)\nonumber\quad,\\
s_+&=&\frac{-7 + \sqrt{73}}{12}\quad .
\end{eqnarray}
$\Gamma_{E}$ is the complete Euler Gamma function, which is defined as
\begin{equation}
\Gamma_{E}(z)=\int_0^{\infty}t^{z-1}e^{-t}dt\quad.
\end{equation}
When $D=0$, Eq.(\ref{finallfR}) should be
consistent with Eq.(\ref{fHubble}) and should go
back to the $\Lambda$CDM model. This puts the
constant in Eq.(\ref{finallfR}) into
$6(1-\Omega_m^0)$. Thus finally we have the
analytic  $f(R)$ form
\begin{equation}
f(R)=[L_{\Lambda}-6(1-\Omega_m^0)](1-\epsilon)+6(1-\Omega_m^0)\label{finallfRwithconstant}\quad .
\end{equation}

\begin{figure}
\includegraphics[width=3.5in,height=3.2in]{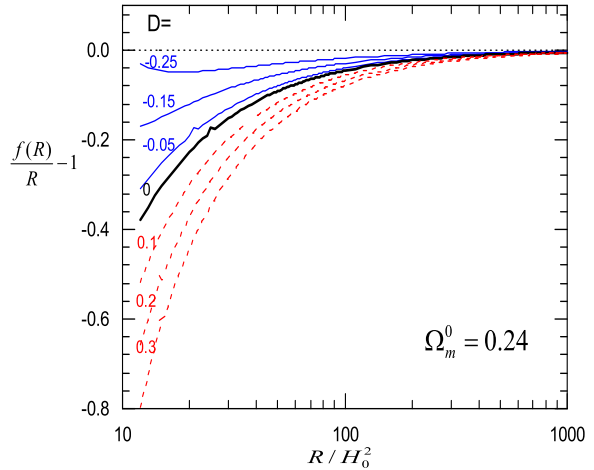}
\caption{$f(R)$ models that have the same expansion history as the $\Lambda$CDM cosmology.}\label{fro}
\end{figure}

Eq.(\ref{finallfRwithconstant}) is a general
analytical expression for $f(R)$ models that can
have the same background expansion history as
that of the $\Lambda$CDM cosmology. The solution
has only two free dimensionless parameters,
namely, $\Omega_m^0$ and $D$.  In order to avoid
the short-timescale instability at high
curvature~\cite{Song}, it requires $D<0$.
Moreover for a viable $f(R)$ model, it also
requires $F>0,f_{RR}>0$ \cite{tsujikawa}. In
Fig~\ref{fro}, we illustrate the $f(R)$ models
obtained from Eq.(\ref{finallfRwithconstant}) and
the blues lines indicate the viable $f(R)$ models which satisfy the condition
mentioned above.

If we do not start from the concrete
model describing the expansion history of the
universe, there is a lot of freedom for choosing
the scalar field $F(x)$ except the boundary
condition Eq.(\ref{boundary}). The $f(R)$ gravity
can be constructed only when the scalar field
$F(x)$ can be reasonably chosen based on physical
motivations. However, in the Jordan frame, there
is no clear physical motivation to provide a
particular form for the field $F(x)$. We need to turn to the Einstein
frame. In the Einstein frame, $F(x)$ relates to a
scalar $\Gamma(x)$ representing the mass dilation
rate $\frac{\partial \ln m}{\partial t}$ in
gravitational field \cite{Tsujikawa,Hefr} and
$\Gamma(x)$ can be deemed as the coupling
strength in the framework of interacting dark
energy models~\cite{7,HJH,Hepertur,coupling}.
$F(x)$ can be determined by constructing $\Gamma$
which, in turn, the $f(R)$ model can be
constructed. In the next section we will shift
our attention to the Einstein frame.

\subsection{The Einstein frame~\label{The Einsteinframe}}
The equations of motion in the Einstein frame can
be obtained by using conformal transformations
\begin{equation}
\tilde{H}=\frac{1}{\sqrt{F}}(1+\frac{1}{2}\frac{d\ln F}{dx})H\quad.
\end{equation}
As shown in~\cite{Hefr}, after the conformal
transformation the expansion history in the
Einstein frame is governed by
\begin{eqnarray}
&&\frac{d\tilde{H}}{d\tilde{x}}=\frac{\tilde{R}}{6\tilde{H}}-2\tilde{H}\label{1}\quad,\\
&&\frac{d\tilde{R}}{d\tilde{x}}+36\Gamma\frac{d\Gamma}{d\tilde{x}}+72\Gamma^2=-3\kappa^2\tilde{\rho}_m(\frac{\Gamma}{\tilde{H}}+1)\label{2}\quad,\\
&&\frac{d\tilde{\rho}_m}{d\tilde{x}}+3\tilde{\rho}_m=\frac{\Gamma}{\tilde{H}}\tilde{\rho}_m\label{3}\quad,
\end{eqnarray}
where $\Gamma$ is the coupling strength between
the scalar field and matter which can also be
explained as the mass dilation rate function
\cite{Hefr}. From the relation between $F$ and
$\Gamma$ and expressing $F$ into $\varphi$, $\ln
F= \kappa \sqrt{\frac{2}{3}}\varphi$, we can
rewrite the continuity equation for the scalar
field into
\begin{equation}
\frac{d\varphi}{d\tilde{x}}=-\frac{\sqrt{6}\Gamma}{\kappa
\tilde{H}}.\label{4}
\end{equation}
Once the dilation function is specified, we can
explore the expansion history of the universe in
the Einstein frame.

$\Gamma$ has clear physical meaning, which
represents the coupling strength of the scalar
field to the matter field. Although $\Gamma$ is a
free field, the choice of $\Gamma$ must be
consistent with the viability condition of the
$f(R)$ gravity. A viable $f(R)$ model must pass
the stringent local test, which requires that
when $\tilde{R}\rightarrow \infty$, the model
should go back to the standard Einstein gravity,
so that $\lim_{\tilde{x}\rightarrow
-\infty}\Gamma=0$.

Eqs.(\ref{1})-(\ref{3})
governing the expansion history are not stable
when $\Gamma<0$ at high curvature region, except
that the coupling strength is extremely small,
for example $|\Gamma|<10^{-10}$. This phenomena
is consistent with that discussed in
~\cite{Ignacy}. Thus a viable $f(R)$ model
requires that  $\Gamma$ is positive
and drops to sufficiently small values in the early
time of the Universe.

$\Gamma$ also is the dilation rate of the mass
of test particles in gravitational fields. To
satisfy the weak equivalent principle, $\Gamma$
should be independent of the species of matter
\cite{Fujii}. One natural choice of the $\Gamma$
form is to consider it as a geometrical quantity,
a function of $\tilde{R}$.  There is quite a wide
range of the choices for $\Gamma$ which can
satisfy the above conditions. In this work, we
only present one of the simplest choices, where
we describe $\Gamma$ as a function of the scalar
Ricci curvature $\tilde{R}$ in the Einstein frame
as
\begin{equation}
\Gamma = \frac{\alpha}{\tilde{R} + \beta}\quad,\label{model}
\end{equation}
where $\alpha$ and $\beta$ are constants.
Eq.(\ref{model}) is just one kind of possible
phenomenological descriptions since we are lack
of the knowledge on the nature of the dilation
rate.

Once we specify the dilation rate $\Gamma$, we
can get the $f(R)$ gravity in the Jordan frame by
using the conformal transformation from the
Einstein frame to the Jordan frame.

\section{perturbation theory in $f(R)$ cosmology \label{sectionperturbedspacetime}}

In this section, we will extend our discussion to the
perturbed space-time.  We will first review the
cosmological perturbation theory in $f(R)$
gravity in the Jordan frame.  We will show that perturbations can be
consistently connected between the Jordan frame and the Einstein frame.

\subsection{The Jordan frame}
We focus on the scalar perturbation. The
perturbed line element in Fourier space reads,
\begin{eqnarray}
ds^2 &=&  a^2[-(1+2\psi Y^{(s)})d\tau^2+2BY^{(s)}_id\tau
dx^i\nonumber \\
&+&(1+2\phi Y^{(s)})\delta_{ij}dx^idx^j+EY^{(s)}_{ij}dx^idx^j]\label{pert_jordan}\quad,
\end{eqnarray}
where $\psi,B,\phi,E$ are scalar fields which
indicate the perturbations. $Y^{(s)}$ is the
scalar harmonic function. The perturbed form of the modified
Einstein equations Eq.(\ref{Einsteineq})
$$\delta {\Sigma_{\nu}}^{\mu}=\kappa^2\delta
{T_{(m)\nu}}^{\mu}$$ can be obtained by writing
\begin{equation}
\delta {\Sigma_{\nu}}^{\mu}=\delta g^{\mu\sigma}\Sigma_{\nu\sigma}+ g^{\mu\sigma}\delta\Sigma_{\nu\sigma}\quad,
\end{equation}
where
\begin{eqnarray}
\delta\Sigma_{\mu\nu}&=&\delta FR_{\mu\nu}+F\delta R_{\mu\nu}-\frac{1}{2}Fg_{\mu\nu}\delta R -\frac{1}{2}f\delta g_{\mu\nu}\nonumber \\
&-&\delta \nabla_{\mu}\nabla_{\nu}F+\delta g_{\mu\nu}\Box F+g_{\mu\nu}\delta\Box F\quad,
\end{eqnarray}
and expressing
\begin{eqnarray}
\delta {T_{(m)0}}^{0}Y^{(s)} & = &-\delta \rho_mY^{(s)}\nonumber\quad, \\
\delta {T_{(m)i}}^{0}Y_i^{(s)} & = &(p_m+\rho_m)(v_m+B)Y_i^{(s)}\nonumber\quad,\\
\delta {T_{(m)0}}^{i}Y^{(s)i} & = & -(p_m+\rho_m)v_mY^{(s)i}\nonumber\quad,\\
\delta {T_{(m)j}}^{i}Y^{(s)i}_j & = &(\delta p_m\delta^i_j+{\Pi_{mj}}^i)Y^{(s)i}_j\quad.
\end{eqnarray}
Inserting the line element Eq.(\ref{pert_jordan})
into the above equations, we can get the
perturbation form of the modified Einstein equation.

From
$\delta {\Sigma_{0}}^{0}=\kappa^2\delta {T_{(m)0}}^{0}$ we obtain,
\begin{eqnarray}
-\frac{\kappa^2}{2}\delta \rho_m a^2&=&F[-k^2\phi+3\mathcal{H}(\mathcal{H}\psi-\phi')+k\mathcal{H}B-\frac{1}{6}k^2E]\nonumber \\
    &+&F'(\frac{k}{2}B+3\mathcal{H}\psi-\frac{3}{2}\phi')-\frac{3}{2}\mathcal{H}^2\delta F\nonumber\\ &+&\frac{3}{2}\frac{a''}{a}\delta F-\frac{3}{2}\mathcal{H}\delta F'-\frac{k^2}{2}\delta F\label{JE_1}\quad,
\end{eqnarray}
where the prime denotes the derivative with
respect to the conformal time $\tau$.

The trace part $\delta {\Sigma_{i}}^{i}=\kappa^2\delta {T_{(m)i}}^{i}$ gives
\begin{eqnarray}
\kappa^2a^2\delta p_m&=&F[-\frac{2}{3}k^2\psi-\frac{2}{3}k^2\phi-\frac{1}{9}k^2E+4\frac{a''}{a}\psi+\frac{2}{3}kB'\nonumber \\
&+&\frac{4}{3}k\mathcal{H}B
   -2\mathcal{H}^2\psi+2\mathcal{H}\psi'-4\mathcal{H}\phi'-2\phi'']\nonumber \\
   &+&F'[\frac{2}{3}kB+\psi'+2\mathcal{H}\psi-2\phi']\nonumber\\
   &+&\delta F[\mathcal{H}^2+\frac{a''}{a}-\frac{2}{3}k^2]-\delta F''-\delta F'\mathcal{H}+2\psi F''\nonumber\quad,\\\label{JE_2}
\end{eqnarray}
and the anisotropic part $\delta
{\Sigma_{i}}^{j}=\kappa^2\delta {T_{(m)i}}^{j}$
($i\neq j$)  gives
\begin{eqnarray}
\kappa^2a^2p_m\Pi_m&=&F[-k^2\psi-k^2\phi+2k\mathcal{H}B+\mathcal{H}E'-\frac{1}{6}k^2E\nonumber\\
&+&\frac{1}{2}E''+kB']+F'(\frac{1}{2}E'+kB)-k^2\delta F.\nonumber\\\label{JE_3}
\end{eqnarray}

The last part $\delta
{\Sigma_{0}}^{i}=\kappa^2\delta {T_{(m)0}}^{i}$
leads to
\begin{eqnarray}
-\frac{1}{2}\kappa^2a^2(\rho_m+p_m)v_m&=&F[k\phi'-k\mathcal{H}\psi+2\mathcal{H}^2B-\frac{a''}{a}B\nonumber \\
&+&\frac{1}{6}kE']+F'[\mathcal{H}B-\frac{1}{2}k\psi]\nonumber \\
&+&\frac{1}{2}k\delta F'-\frac{1}{2}k\mathcal{H}\delta F-\frac{1}{2}B F''.\nonumber\\\label{JE_4}
\end{eqnarray}

The above perturbation equations are covariant.
This can be easily seen by investigating the
infinitesimal coordinate transformation,
\begin{eqnarray}
\hat{x^{\mu}}& = &x^{\mu}+\delta x^{\mu}\nonumber\quad, \\
\delta x^{0} & = & \xi^0 Y^{(s)}\nonumber\quad,\\
\delta x^{i} & = & \beta Y^{(s)i}\nonumber\quad.
\end{eqnarray}
We use the symbol  with ``hat" to indicate the
quantities after coordinate transformation. Under
the infinitesimal transformation, we can show
that the perturbed quantities in the line element
Eq.~(\ref{pert_jordan}) behave as~\cite{Hepertur}
\begin{eqnarray}
\hat{\psi}Y^{(s)}&=& (\psi-{\xi^0}'-\mathcal{H}\xi^0)Y^{(s)}\nonumber\quad,\\
\hat{\phi}Y^{(s)}&=& (\phi-\frac{1}{3}k\beta-\mathcal{H}\xi^0)Y^{(s)}\nonumber \quad,\\
\hat{B}Y^{(s)}_i&=&(B-k\xi^0-\beta')Y^{(s)}_i\nonumber\quad, \\
\hat{E}Y^{(s)}_{ij}&=&(E+2k\beta)Y^{(s)}_{ij}\quad.\label{transJordan}
\end{eqnarray}
After inserting Eq. (\ref{transJordan}) into Eqs.
(\ref{JE_1})-(\ref{JE_4}), we find that Eqs.
(\ref{JE_1})-(\ref{JE_4})  keep the same forms
under the infinitesimal transformation, which
shows that Eqs. (\ref{JE_1})-(\ref{JE_4}) are
covariant. Furthermore, if we take $F\rightarrow
1$,$\delta F\rightarrow 0$, Eqs.
(\ref{JE_1})-(\ref{JE_4}) can go back to the
standard cosmological perturbation equations in the
Einstein gravity as presented in ~\cite{Kodama}.

After we get the perturbed form of the modified Einstein
equations, we turn to derive the equation of
motion for matter field in the Jordan frame. From
\begin{equation}
\nabla^{\nu}T_{\mu\nu}^{(m)}=0\label{eqofmotion}
\end{equation}
we can obtain its perturbed form  $\delta
\nabla^{\nu}T_{\mu\nu}^{(m)}=0 $.  Looking at the
zero-th component, we have
\begin{eqnarray}
&&\delta_{m}'+3\mathcal{H}(\frac{\delta p_{m}}{\delta
\rho_{m}}-w_{m})\delta_{m}\nonumber \\
&=&-(1+w_{m})kv_{m}
-3(1+w_{m})\phi'\quad.\label{EQM}
\end{eqnarray}
The i-th component gives
\begin{eqnarray}
&&\left[(\rho_{m}+p_{m})(B+v_{m})\right
]'+4\mathcal{H}(p_{m}+\rho_{m})(B+v_{m}) \nonumber \\
&=& (p_{m}+\rho_{m})k\psi+k\delta
p_{m}-\frac{2}{3}kp_m\Pi_m\quad.\label{EQV}
\end{eqnarray}
Similarly, it can be shown that the above
equations are covariant under the infinitesimal
coordinate transformation.

The freedom of choosing coordinates in general
relativity will lead to ambiguity in defining a
real density perturbation from the coordinate
transformation. The systematic way to get rid of
the ambiguity is to fix the gauge, which has been
well discussed in \cite{Kodama}. Here we will
argue that the gauge conditions in the Jordan
frame are not always the same as the conditions
in the Einstein frame after the conformal
transformation.

We only focus our discussion on the most commonly
used gauges, the Newtonian gauge and the
Synchronous gauge. The Newtonian gauge is defined
by setting $B=E=0$ and these conditions
completely fix the gauge
\begin{eqnarray}
\xi^0 &=& -\frac{\hat{B}}{k}-\frac{\hat{E}'}{2k^2}\nonumber\quad, \\
\beta &=& \frac{\hat{E}}{2k}\quad.
\end{eqnarray}
The perturbations in this gauge can be shown as
gauge invariant. For instance, the matter density
perturbation can be presented as
\begin{equation}
\delta^{(N)}=\hat{\delta}-\frac{\rho_m'}{\rho_m}(\frac{\hat{B}}{k}+\frac{\hat{E}'}{2k^2})\quad,
\end{equation}
where $\hat{\delta}$ is a general perturbation.
We can show that $\delta^{(N)}$ is invariant
under the coordinate transformation Eq.(
\ref{transJordan}).

However, in contrast to the Newtonian gauge, the
Synchronous gauge is not completely fixed because
the gauge condition $\psi=B=0$ only confines the
gauge up to two arbitrary constants $C_1,C_2$,
\begin{eqnarray}
\xi^0 &=& \frac{C_1}{a}\nonumber\quad, \\
\beta &=& -kC_1\int\frac{d\tau}{a}+C_2\quad.
\end{eqnarray}
In order to fix the Synchronous gauge, additional
conditions are called for. Usually, $C_2$ is
fixed by specifying the initial condition for the
curvature perturbation in the early time of the
universe and $C_1$ can be fixed by setting the
peculiar velocity of DM to be zero, $v_m=0$,
which is equivalent to say that the gauge is at
rest with respect to the cold DM and the observer
is comoving with cold DM. After fixing $C_1,C_2$,
the Synchronous gauge can be completely fixed.

As long as the Synchronous gauge is completely
fixed, it will be equivalent to the Newtonian
gauge. For instances, the density perturbations
in different gauges are related by
\begin{equation}
\delta^{(N)}=\delta^{(S)}+\frac{\rho_m'}{\rho_m}\frac{v^{N}-v^{S}}{k}\quad,
\end{equation}
where $v^{N}=\hat{v}-\frac{\hat{E}'}{2k}$,$\quad
v^{S}=0$.

We will show later that the gauge condition for
the Newtonian gauge will be the same both in the
Jordan frame and the Einstein frame. However we
will see that the gauge condition for the
Synchronous gauge will change when we transform
the Jordan frame to the Einstein frame by using
the conformal transformation.

Since the Newtonian
gauge has this advantage, we will work in the Newtonian gauge
hereafter. We use the Bardeen potentials
~\cite{Bardeen} $\Phi=\phi$, $\Psi=\psi$ to
represent the space time perturbations. Since we
are only interested in the DM dominated period in
the  $f(R)$ cosmology, we  set $p_m=0$ and
$\delta p_m=0$. The perturbed Einstein equations
can be reduced to
\begin{eqnarray}
-\frac{\kappa^2}{2}\delta \rho_m a^2&=&F[-k^2\Phi+3\mathcal{H}(\mathcal{H}\Psi-\Phi')]\nonumber \\
    &+&F'(3\mathcal{H}\Psi-\frac{3}{2}\Phi')-\frac{3}{2}\mathcal{H}^2\delta F\nonumber\\ &+&\frac{3}{2}\frac{a''}{a}\delta F-\frac{3}{2}\mathcal{H}\delta F'-\frac{k^2}{2}\delta F\label{JEN_1}\quad ,\\
\delta F''&=&-\mathcal{H}\delta F'+F[-\frac{2}{3}k^2\Psi-\frac{2}{3}k^2\Phi+4\frac{a''}{a}\Psi\nonumber \\
&-&2\mathcal{H}^2\Psi+2\mathcal{H}\Psi'-4\mathcal{H}\Phi'-2\Phi'']+2\Psi F''\nonumber \\
   &+&F'[\Psi'+2\mathcal{H}\Psi-2\Phi']\nonumber \\
   &+&\delta F[\mathcal{H}^2+\frac{a''}{a}-\frac{2}{3}k^2]\label{JEN_2}\quad ,\\
\frac{\delta F}{F}&=&-\Psi-\Phi\label{JEN_3} \quad , \\
-\frac{1}{2}\kappa^2a^2\rho_mv_m&=&F[k\Phi'-k\mathcal{H}\Psi]-\frac{1}{2}kF'\Psi\nonumber \\
&+&\frac{1}{2}k\delta F'-\frac{1}{2}k\mathcal{H}\delta F\quad.\label{JEN_4}
\end{eqnarray}

$F$ is a free scalar field which characterizes
the deviation from the Einstein gravity. The
perturbation $\delta F$ can be obtained by
\begin{equation}
\delta F=\frac{\partial F}{\partial R}\delta R \quad,\label{delta_F_J}
\end{equation}
where
\begin{equation}
a^2\delta R=\mathcal{H}(18\Phi'-6\Psi')+4k^2\Phi+2k^2\Psi-12\Psi\frac{a^{''}}{a}+6\Phi^{''}\quad.\label{RJ}
\end{equation}

Using Eq.(\ref{RJ}) and Eq.(\ref{JEN_1}) to
eliminate $\Phi^{''}$  and  $k^2\Phi$ in
Eq.(\ref{JEN_2}), we obtain
\begin{eqnarray}
&&\delta F''+2\mathcal{H}\delta F'+a^2(\frac{k^2}{a^2}+M^2)\delta F\nonumber\\
 &=&\frac{\kappa^2}{3}a^2\delta \rho_m +F'(4\mathcal{H}\Psi-3\Phi'+\Psi')+2\Psi F''\quad,\label{deltF}
\end{eqnarray}
where
\begin{equation}
M^2=\frac{1}{3}(\frac{F}{f_{RR}}-R)\label{Mdef}\quad,
\end{equation}
and
\begin{equation}
R=\frac{6a''}{a^3}\quad,\quad f_{RR}=\frac{\partial F}{\partial R}\quad.
\end{equation}
Eq.(\ref{deltF}) is consistent with Eq.(8.92) in
~\cite{tsujikawa}, if we express it in the cosmic
time $t$.

>From the equations of motion Eqs.(\ref{EQM})
(\ref{EQV}), we obtain
\begin{eqnarray}
&&\delta_m'=-kv_m-3\Phi'\label{mp_jordan}\quad ,\\
&&v_m'+\mathcal{H}v_m=k\Psi\label{v_jordan}\quad .
\end{eqnarray}
Combining Eqs.(\ref{mp_jordan},~\ref{v_jordan}),
we have the second order differential equation for the
evolution of the matter perturbation
\begin{equation}
\delta_m^{''}+\mathcal{H}\delta_m'+k^2\Psi=-3\mathcal{H}\Phi'-3\Phi''\label{secondmE}\quad.
\end{equation}

\subsection{The Einstein frame}

We will first derive the perturbed form of the equation of
motion for matter fields Eq.(\ref{eqofmotion})
under the general conformal transformation. Then
we will fix the freedom of the general conformal
transformation by setting $\Omega^2=F$ to obtain
the perturbation form of the modified Einstein equations. We
will show that the equations obtained under
conformal transformation from the Jordan frame
are consistent with the results derived by doing
perturbations directly in the Einstein frame.

In the background, the conformal transformation
is defined by re-scaling the line element of the
Jordan frame
\begin{equation}
\tilde{ds}^2=\Omega^2ds^2 \quad .
\end{equation}
However, in perturbed spacetime, we need to take account of the perturbation of $\Omega$. And thus the full line element reads
\begin{eqnarray}
\tilde{ds}^2&=& (\Omega+\delta\Omega)^2ds^2\nonumber \\
&=&  \tilde{a}^2[-(1+2\tilde{\psi})d\tau^2+2\tilde{B}d\tau
dx^i\nonumber \\
&+&(1+2\tilde{\phi})\delta_{ij}dx^idx^j+\tilde{E}dx^idx^j]\quad .\label{perturbedspacetime}
\end{eqnarray}
Since we are only interested in the linear
perturbation, we can expand the above expression
to the linear order
\begin{eqnarray}
\psi&=&\tilde{\psi}-\delta \ln\Omega\nonumber\quad, \\
\phi&=&\tilde{\phi}-\delta \ln \Omega\nonumber\quad,\\
B&=&\tilde{B}\nonumber\quad,\\
E&=&\tilde{E}\quad,\label{trans}
\end{eqnarray}
where the symbols with ``tilde" indicate the
quantities in the Einstein frame.

For the clear discussion, hereafter we will use
``tilde" to indicate quantities that  are
different in the Einstein frame from their values
 in the Jordan frame. For the quantities without
``tilde", they are the same in both frames. We
will use ``hat" to indicate the coordinate
transformation which should not be confused with
the conformal transformation. The differential
operator $d$ does not depend on the geometric
structure $g_{ab}$ on the spacetime manifold, it
will not change under the conformal
transformation so that the ordinary derivative
operator $\partial_a$ and the conformal
coordinates $d\tau, dx$ will be the same in both
the Einstein frame and the Jordan frame.

Noting
\begin{equation}
\hat{\delta \ln\Omega}=\delta \ln\Omega -(\ln\Omega)'\xi^0\quad,
\end{equation}
and considering
Eqs.(\ref{trans},\ref{transJordan}), we can show
that, under the infinitesimal coordinate
transformation, the perturbed quantities
$\tilde{\psi}$, $\tilde{\phi}$ behave as
\begin{eqnarray}
\hat{\tilde{\psi}}&=& \tilde{\psi}-{\xi^0}'-\mathcal{\tilde{H}}\xi^0\nonumber\quad,\\
\hat{\tilde{\phi}}&=& \tilde{\phi}-\frac{1}{3}k\beta-\mathcal{\tilde{H}}\xi^0\quad ,
\end{eqnarray}
where
\begin{equation}
\mathcal{H}=\tilde{\mathcal{H}}-(\rm ln \Omega)'\quad.
\end{equation}
It is clear that in the coordinate transformation
$\tilde{\psi}$, $\tilde{\phi}$ behave in a
similar way as in the Jordan frame. This point
also holds for other perturbation quantities
\begin{eqnarray}
\hat{B}&=&B-k\xi^0-\beta'\nonumber \quad,\\
\hat{E}&=&E+2k\beta\nonumber\quad.
\end{eqnarray}

Now we can discuss the gauge conditions
in the Einstein frame. In the Newtonian gauge, the  gauge conditions
$B=E=0$ fix the gauge as that in the Jordan frame, namely
\begin{eqnarray}
\xi^0 &=& -\frac{\hat{B}}{k}-\frac{\hat{E}'}{2k^2}\nonumber\quad, \\
\beta &=& \frac{\hat{E}}{2k}\quad .
\end{eqnarray}
Thus the Newtonian gauge conditions keep the same under the conformal transformation.

However, in the Synchronous gauge, after the
conformal transformation, the gauge conditions in
the Einstein frame read
\begin{eqnarray}
\tilde{\psi}&=&\delta \ln\Omega\nonumber \quad,\\
B&=&0\nonumber\quad,\\
v_m&=&0\quad .
\end{eqnarray}
We will find that $v_m=0$ is a physical choice
because it is the solution of the equation of
motion in Eq.(\ref{v_Ein}). It is clear that in
the Einstein frame the gauge conditions for the
Synchronous gauge are not exactly the same as
those in the Jordan frame.

Considering the gauge conditions for the Newtonian
gauge are the same in both the Einstein frame and
the Jordan frame, we will work in Newtonian gauge
hereafter.

In the background, the conformal transformation
leads the equation of motion of the matter field
Eq.(\ref{eqofmotion}) in the Einstein frame to
the form
\begin{equation}
\tilde{\nabla}_{\mu}\tilde{T}^{\mu \nu}=-\tilde{T}\frac{\tilde{\partial}^{\nu}\Omega}{\Omega}\quad \label{efm}\quad ,
\end{equation}
where $\tilde{T}^{\mu
\nu}\equiv\tilde{g}^{\mu\tau}\tilde{g}^{\nu\sigma}\tilde{T}_{\tau
\sigma}$, $T^{\mu \nu}\equiv
g^{\mu\tau}g^{\nu\sigma}T_{\tau \sigma}$ and
\begin{equation}
\tilde{T}_{\mu\nu}=\frac{1}{\Omega^2}T_{\mu\nu}\quad,\tilde{T}^{\mu\nu}=\frac{1}{\Omega^6}T^{\mu\nu}\quad,\tilde{T}=\frac{T}{\Omega^4}.\label{tensortrans}
\end{equation}
$-\tilde{T}\frac{\tilde{\partial}^{\nu}\Omega}{\Omega}$
is the coupling vector. We only consider the
evolving universe where the coupling vector is
time-like. We focus on the DM dominated phase
with $p_m=0$.

The conservation equation  Eq.(\ref{efm}) can be
recasted into
\begin{equation}
\tilde{U}^0\frac{\partial\tilde{\rho}_m}{\partial \tau}+3\tilde{H}\tilde{\rho}_m=\Gamma\tilde{ \rho}_m\quad , \label{matter}
\end{equation}
where $\tilde{U}^0=1/\tilde{a}$, $\Gamma$ is the
mass dilation rate
\begin{equation}
\Gamma=-\tilde{U}^0\frac{\partial\ln \Omega}{\partial \tau}=\frac{1}{\tilde{m}}\tilde{U}^0\frac{\partial\tilde{m}}{\partial\tau}. \label{dilation}
\end{equation}

The perturbed equation of Eq.(\ref{matter}) reads
\begin{equation}
\tilde{\delta}_m'=-kv_m-3\tilde{\Phi}'+\tilde{a}\Gamma\tilde{\Psi}+\tilde{a}\delta\Gamma\label{permatter}\quad,
\end{equation}
where we have used
\begin{eqnarray}
&&3\delta \tilde{H}=\frac{1}{\tilde{a}}kv_m+\frac{3}{\tilde{a}}(\tilde{\Phi}'-\mathcal{\tilde{H}}\tilde{\Psi})\nonumber\quad,\\
&&\delta \Gamma = \tilde{\Psi}\tilde{U}^0(\ln \Omega)' -\tilde{U}^0(\delta \ln \Omega)'\label{deltaH}
\end{eqnarray}
and $\delta \tilde{U}^0 =-\tilde{\psi}\tilde{U}^0$.

In the Einstein frame, the perturbation equation for the peculiar
velocity has the form
\begin{equation}
v_m'+\mathcal{\tilde{H}}v_m=\tilde{a}\Gamma(\tilde{v}_t -v_m)+k\tilde{\Psi}\label{v_Ein}\quad,
\end{equation}
where
\begin{equation}
\tilde{v}_t =k\frac{\delta \ln \Omega}{(\ln \Omega)'}\quad.
\end{equation}
Eqs.(\ref{permatter}, \ref{v_Ein}) are consistent
with the results in~\cite{199} and~\cite{7} by
setting the general coupling vector as
$\tilde{Q}^{\mu}=\Gamma \tilde{\rho}_m
\tilde{U}^{\mu}_t$, where $\tilde{U}^{\mu}_t$ is
along the direction of
$\tilde{\partial}^{\nu}\Omega\quad$.

Eqs.(\ref{permatter}, \ref{v_Ein}) can also be
consistently obtained from the perturbation
equations of Eqs.(\ref{mp_jordan},
\ref{v_jordan}) by doing the
conformal transformation
\begin{eqnarray}
&&a=\frac{\tilde{a}}{\Omega}\nonumber \quad,\\
&&v_m=\tilde{v}_m\nonumber \quad,\\
&&U^0=\Omega \tilde{U}^0\nonumber\quad, \\
&&\rho_m=\Omega^4\tilde{\rho}_m \nonumber\quad, \\
&&\delta \rho_m = \delta \Omega^4 \tilde{\rho}_m + \Omega^4\delta \tilde{\rho}_m\label{contrans}\quad, \\
&&\delta_m=\tilde{\delta}_m+4(\delta \ln \Omega)\quad,\nonumber\\
&&\Psi=\tilde{\Psi}-\delta \ln\Omega\nonumber\quad, \\
&&\Phi=\tilde{\Phi}-\delta \ln \Omega\nonumber\quad.
\end{eqnarray}

Similarly, combining Eqs.(\ref{permatter},
\ref{v_Ein}), we can get the second order
differential equation
\begin{eqnarray}
&&\tilde{\delta}_m^{''}+(\mathcal{\tilde{H}}+\tilde{a}\Gamma)\tilde{\delta}_m'+k^2\tilde{\psi}\nonumber \\
&=&(\mathcal{\tilde{H}}+\tilde{a}\Gamma)\tilde{a}\delta \Gamma-k\tilde{a}\Gamma\tilde{v}_t+(\tilde{a}\delta \Gamma)'\nonumber\\
&+&(\mathcal{\tilde{H}}+\tilde{a}\Gamma)(\tilde{a}\Gamma\tilde{\psi}-3\tilde{\phi}')-3\tilde{\phi}''+(\tilde{a}\Gamma\tilde{\psi})'\quad.
\end{eqnarray}

Next, we consider the perturbed Einstein
equations in the Einstein frame. We set
$\Omega^2=F$ to fix the freedom of the general
conformal transformations. Inserting the
conformal transformations Eq.(\ref{contrans})
into Eqs.(\ref{JEN_1})-(\ref{JEN_4}), we obtain
\begin{eqnarray}
-\frac{\kappa^2}{2}(\delta \tilde{\rho}_m+\delta \tilde{\rho}_d) \tilde{a}^2&=&-k^2\tilde{\Phi}+3\mathcal{\tilde{H}}(\mathcal{\tilde{H}}\tilde{\Psi}-\tilde{\Phi}')\quad ,\\
\kappa^2\tilde{a}^2\delta\tilde{p}_d&=&-\frac{2}{3}k^2\tilde{\Psi}-\frac{2}{3}k^2\tilde{\Phi}+4\frac{\tilde{a}''}{\tilde{a}}\tilde{\Psi}\nonumber \\
&-&2\mathcal{\tilde{H}}^2\tilde{\Psi}+2\mathcal{\tilde{H}}\tilde{\Psi}'-4\mathcal{\tilde{H}}\tilde{\Phi}'-2\tilde{\Phi}''\quad ,\nonumber\\
0&=&-\tilde{\Psi}-\tilde{\Phi}\nonumber \quad , \\
k\tilde{\Phi}'-k\mathcal{\tilde{H}}\tilde{\Psi}&=&-\frac{1}{2}\kappa^2\tilde{a}^2[\tilde{\rho}_mv_m+(\tilde{\rho}_d+\tilde{p}_d)\tilde{v}_d]\nonumber\quad ,
\end{eqnarray}
where $\tilde{v}_d$, $\tilde{\rho}_d$ and $\tilde{p}_d$ are defined as
\begin{eqnarray}
\tilde{v}_d&=&-k\frac{\delta \ln F}{2\tilde{a}\Gamma}\nonumber\quad, \\
\tilde{\rho}_d&=&\frac{3\Gamma^2}{\kappa^2}+V\nonumber\quad, \\
\tilde{p}_d&=&\frac{3\Gamma^2}{\kappa^2}-V\nonumber\quad, \\
V&=&\frac{FR-f}{2\kappa^2F^2}\quad .\label{def}
\end{eqnarray}
The perturbed Einstein equations give rise to the perturbation of $\delta V$ as
\begin{equation}
\kappa^2\tilde{a}^2\delta V=(\delta \ln F)(9\tilde{a}'\Gamma+2\mathcal{\tilde{H}}^2-3\tilde{a}^2\Gamma^2-\frac{\tilde{a}''}{\tilde{a}}+3\tilde{a}\Gamma')\quad.\label{deltaV}
\end{equation}
where we have used
\begin{equation}
\delta \Gamma=-\tilde{\Psi}\Gamma-\frac{1}{2\tilde{a}}(\delta \ln F)'\label{deltagamma}\quad .
\end{equation}
If we directly perturb this $V$ expressed in Eq.(\ref{def})
\begin{equation}
\kappa^2\delta V=\frac{1}{2}\frac{\delta F R}{F^2}-\frac{(FR-f)\delta F}{F^3}\quad,
\end{equation}
after inserting
\begin{equation}
R=6F(3\frac{\mathcal{\tilde{H}}}{\tilde{a}}\Gamma+\Gamma^2+\frac{\tilde{a}''}{\tilde{a}^3}+\frac{\Gamma'}{\tilde{a}})\label{RJE}\quad,
\end{equation}
eliminating $f$  by the Friedmann
equation and further employing
\begin{equation}
\kappa^2\tilde{\rho}_m=4\frac{\mathcal{\tilde{H}}^2}{\tilde{a}^2}-2\frac{\tilde{a}''}{\tilde{a}^3}-6\Gamma^2\quad,
\end{equation} we can get the result in consistent with
Eq.(\ref{deltaV}).

$\delta F$ can be determined by Eq.(\ref{deltF}).
Inserting Eq.(\ref{contrans}), (\ref{deltF})
turns out to be
\begin{eqnarray}
\delta F''&=&[-k^2+\frac{2}{3}\kappa^2\tilde{a}^2\tilde{\rho}_m-2\tilde{a}^2\Gamma^2+2\frac{\tilde{a}''}{\tilde{a}}\nonumber\\
&+&4\tilde{a}\Gamma'+12\tilde{a}'\Gamma-\frac{1}{3}\frac{\tilde{a}^2}{f_{RR}}]\delta F+\frac{1}{3}\kappa^2\tilde{a}^2F\delta\tilde{\rho}_m\nonumber \\
&-&(2\mathcal{\tilde{H}}+4\tilde{a}\Gamma)\delta F'-2\tilde{a}F\Gamma\tilde{\Psi}'-4F\tilde{a}\Psi\Gamma'\nonumber\\
&-&12F\Psi\tilde{a}'\Gamma+6F\tilde{a}\tilde{\Phi}'\Gamma\quad ,
\end{eqnarray}
where $f_{RR}$ is given by
\begin{equation}
f_{RR}=\frac{F'}{R'}=\frac{2\tilde{a}F\Gamma}{R'}\quad ,
\end{equation}
and $R$ can be obtained from Eq.(\ref{RJE}) in
the Einstein frame.

After fixing $\delta F$, we can obtain the
consistent covariant perturbation for $\Gamma$ by
Eq.(\ref{deltagamma}), since the infinitesimal
coordinate transformation for $\delta \Gamma$
presented in Eq.(\ref{deltagamma}) satisfies the
rules of the covariant transformation for a
scalar field
\begin{equation}
\delta \hat{\Gamma}=\delta \Gamma -\Gamma'\xi^0\quad.
\end{equation}

Similarly the perturbation for $H$, $\delta H$ is
given by its covariant configuration
\begin{equation}
3H=\nabla_{\mu}U^{\mu}\quad,
\end{equation}
and so does $\delta \Gamma$
\begin{equation}
\Gamma = -\tilde{U}^{\mu}\tilde{\nabla}_{\mu}\ln \Omega\quad.
\end{equation}

\section{Subhorizion approximation\label{subhoizion}}
\begin{figure}
\includegraphics[width=3.2in,height=2.8in]{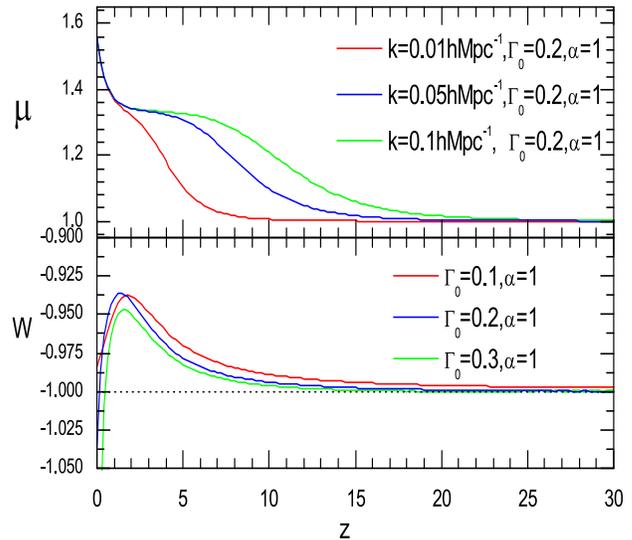}
\caption{The evolutions of $\mu$ and the effective
DE equation of state $w$  in the Jordan frame
for the $f(R)$
models we constructed.
}\label{evolution}
\end{figure}

From the above analysis, $f(R)$ model in the
Jordan frame is conformally equivalent to the
interaction model in the Einstein frame.  This
equivalence holds not only in the background
dynamics but also in the perturbed spacetime.
Therefore, we can construct $f(R)$ models by
choosing a peculiar form of the mass dilation
rate function $\Gamma$ in the Einstein frame. One
of the simplest choice of the $\Gamma$ is $$
\Gamma = \frac{\alpha}{\tilde{R} + \beta}\quad
.$$ After fixing the mass dilation rate function,
we can figure out the evolution of the
cosmological model in the Einstein frame and then
we can transform it into the Jordan frame by
using the conformal transformation. In the lower
panel of Fig~\ref{evolution}, we present some
examples for the effective DE equation of state
$w$ for the $f(R)$ cosmology in the Jordan frame
constructed by selecting the above mass dilation
rate function in the Einstein frame after the
conformal transformation. We can see that the
effective DE equation of state in the constructed
$f(R)$ model presents the phenomena in the
vicinity of $-1$ and with the $-1$ crossing
behavior, which is consistent with present
cosmological observations. This shows that with
the appropriate choice of the mass dilation
function, which refers to the mass, a
more fundamental quantity,  we can construct the
viable $f(R)$ model to  produce the
observationally allowed effective DE equation of
state.

We have also shown that the linear perturbation
equations are exactly equivalent at all scales in
both the Einstein frame and the Jordan frame. In
the usual $f(R)$ model, it was argued that the
scale factor  described by the $f(R)$ cosmology
during the matter phase behaves inconsistently
with cosmological observations\cite{AmendolaPRL}.
It is of interest to examine the $f(R)$ model
constructed from the peculiar mass dilation rate
function we have chosen and see whether our
$f(R)$ model is viable if comparing with
observations in the matter dominated phase. We
assume that the quantities defined in the Jordan
frame are consistent with the actual
observational quantities. We will study the
subhorizion approximation in the Jordan frame.

Following~\cite{tsujikawa}, from
Eqs.(\ref{JEN_1},\ref{JEN_3}), we obtain
\begin{eqnarray}
\Phi &\thickapprox& -\frac{1}{2}\delta \ln F + \frac{a^2}{k^2}\frac{\kappa^2}{2F}\delta \rho_m\nonumber\quad,\\
\Psi &\thickapprox& -\frac{1}{2}\delta \ln F - \frac{a^2}{k^2}\frac{\kappa^2}{2F}\delta \rho_m\quad ,
\end{eqnarray}
where $\delta F$ is governed by Eq.(\ref{deltF})
\begin{equation}
\frac{\delta \ln F}{\delta \rho_m}\approx\frac{\kappa^2}{3F}\frac{1}{\frac{k^2}{a^2}+M^2}\quad,
\end{equation}
and $M^2$ is defined by Eq.(\ref{Mdef}). From
Eq.(\ref{secondmE}), the equation for the growth
function $G(z,k)=\frac{d\ln D}{dx}$ in the
sub-horizon approximation reads
\begin{equation}
\frac{dG}{dx}+G^2+(2+\frac{d\ln H}{dx})G-\frac{3}{2}\Omega_m\mu(k,x)=0\label{growthfunction}\quad ,
\end{equation}
where $D(z,k)=\frac{\delta_m(z,k)}{\delta_m(z_i,k)}$ and
\begin{equation}
\mu(k,x) = \frac{1}{F}\frac{4+3M^2a(x)^2/k^2}{3(1+M^2a(x)^2/k^2)}\quad.
\end{equation}
The  $M^2$ can be very large in the early time of
the universe, since $\Gamma \rightarrow 0$ as
$R\rightarrow \infty$. When $M^2 \gg
\frac{a^2}{k^2}$, the regime is called ``General
Relativistic(GR) regime" where the matter density
perturbation has the standard evolution since the
modified gravity  doesn't show up in this regime
where $\delta F \rightarrow 0$ and
$\Phi/\Psi\rightarrow-1$, $\mu(k,x)\rightarrow1$.
However, in the late time of the Universe, when
$M^2 \ll \frac{a^2}{k^2}$, the modified gravity
effect has prominent influence on the structure
formation because the ratio of
$\Phi/\Psi\rightarrow-1/2$, which has been
altered greatly from the standard Einstein model
and in this case $\mu(k,x)\rightarrow
\frac{4}{3F_0}$ where $F_0$ is the value of $F$
today. This regime is called the
``scalar-tensor(ST) regime" and obviously matter
perturbation does not have the standard evolution
as in the Einstein gravity in this regime. As
shown in the upper panel of Fig~\ref{evolution},
in the transition phase of these two regimes,
$\mu(k,x)$ is strongly scale dependent on the
wave number of $k$. The transition of these two
regimes can be characterized by $ M^2 = k^2/a^2$
which can occur as early as in the matter
dominated epoch, even in small scales, in certain
types of $f(R)$  models in the expansion history
of the Universe~\cite{AmendolaPRL}. In our model,
since the dilation rate function $\Gamma$ will
drop rapidly in the early time of the Universe,
the transition usually happens very late $z<30$
for interested wave number $k$ . Thus in our
constructed $f(R)$ models, we have the standard
matter dominated phase and matter power spectrum,
which can avoid the problems for the usual $f(R)$
model mentioned in \cite{AmendolaPRL}.

\section{conclusions\label{conclusions}}
In this work, we have further disclosed the
equivalence of the $f(R)$ gravity presented in
the Jordan frame and in the Einstein frame from
the background dynamics to the perturbation
theory. In the Jordan frame we have derived the
general analytic solution for the $f(R)$ model to
have the same expansion history as the
$\Lambda$CDM cosmology. This analytic solution of
the $f(R)$ model can provide further insights on
how cosmology can be used to test the gravity at
the largest scales. Moreover we have presented a
systematic and self-consistent way to construct
the $f(R)$ model in the Jordan frame through
conformal transformation by using the mass
dilation rate function in the Einstein frame. We
have also developed linear perturbation theories
to further exhibit the equivalence of the $f(R)$
gravity in the Jordan frame and in the Einstein
frame. We have shown that the $f(R)$ model
constructed from the mass dilation rate can
produce observational allowed effective DE
equation of state and provide the matter phase in
consistent with cosmological observations. It is
of great interest to confront our $f(R)$ model
with latest observations in the future study.

It is interesting to note that our $f(R)$ model
was constructed from the mass dilation rate,
which refers to a fundamental quantity, the mass,
rather than the mysterious DE equation of state.
This gives the root for understanding the
equivalence between the $f(R)$ model in the
Jordan frame and the Einstein frame, since we know well that
the concept of mass plays a central role in
gravitational theories ~\cite{Brans_Dicke,dicke}.
The mass can be classified into the inertial mass
and the gravitational mass. In the Jordan frame
the $f(R)$ gravity can be effectively treated with the modified
gravitational constant $G_{eff}=\frac{G}{F}$ if
we compare it with the Einstein gravity. This is
effectively equivalent to rescale the
gravitational mass $\tilde{m}_g=m\sqrt{F}$.
However the inertial mass in the Jordan frame is
conserved so that the equation of motion for a
free particle in the Jordan frame is described by
$u^{\mu}{\nabla}_{\mu}p^{\nu}=0$, which is the
same as the description in the standard Einstein
gravity, where the rest inertial mass of particle
is conserved.

Any change in the gravitational mass in the
Jordan frame inevitably changes the gravitational
field, which, in turn, will change the inertial
frame. Any change in the inertial frame will
induce the well-known ``frame-dragging" effect
which was recently confirmed by the Gravity Probe
B mission~\cite{framedragging}. The inertial
``frame-dragging" effect can be equivalently
considered as the ``inertial mass-dragging"
effect by assuming that the inertial frame is
unchanged while  the inertial mass of particles
defined in this frame is rescaled. This
understanding can help us interpret that after
the conformal transformation, in the Einstein
frame, the equation of motion for particles in
the inertial frame is governed by the varying
mass equation of motion $
\tilde{u}^{\mu}{\tilde{\nabla}}_{\mu}\tilde{p}^{\nu}=\tilde{m}\tilde{\partial}^{\mu}\ln
\sqrt{F}=\frac{d\tilde{m}}{d\tilde{t}}(\frac{\partial}{\partial\tilde{t}})^{\nu},
$ which is one of the well established physics in
General Relativity~\cite{Ackeret,Hefr}. The rest
inertial mass in the Einstein frame is indeed
rescaled as $\tilde{m}_I=m\frac{1}{\sqrt{F}}$ and
is no longer conserved.

Considering the Mach principle, there should be
no difference among  gravitational theories
described in different frames. Thus a viable
$f(R)$ model in the Jordan frame should be
consistent with the viable model in the Einstein
frame in describing cosmology. The result
obtained in the work has solid physics root.

\emph{Acknowledgment: J.H.He would like to thank
B. R. Granett and L. Guzzo for helpful
discussions. J.H.He acknowledges the Financial
support of MIUR through PRIN 2008 and ASI through
contract Euclid-NIS I/039/10/0. The work of
B.Wang was partially supported by NNSF of China
under grant 10878001 and the National Basic
Research Program of China under grant
2010CB833000.}

\end{document}